\documentclass[aps,prl,preprint,onecolumn,amsmath,amssymb,groupedaddress]{revtex4-2}
\usepackage{graphicx}
\usepackage{chemformula} 
\usepackage{bm}
\usepackage{physics}
\usepackage{upgreek}
\usepackage{titlesec}
\usepackage{hyperref}
\usepackage{xfrac}
\usepackage{amsmath}

\titleformat{\section}[hang]{\normalfont\large\bfseries}{\thesection}{1em}{}
\titleformat{\subsection}[hang]{\normalfont\normalsize\bfseries}{\thesubsection}{1em}{}
\titleformat{\subsubsection}[hang]{\normalfont\normalsize\itshape}{\thesubsubsection}{1em}{}

\begin{document}

\title{Meso-chiral optical properties of plasmonic nanoparticles: uncovering hidden chirality\\}

\author{Yuanyang Xie*}
\author{Alexey V. Krasavin}
\author{Anatoly V. Zayats*}

\affiliation{Department of Physics and London Centre for Nanotechnology, King's College London, London, WS2R 2LS, UK \\
*Email Address: yuanyang.xie@kcl.ac.uk, a.zayats@kcl.ac.uk.}

\date{\today}
\clearpage
\begin{abstract}
Molecular chirality plays an important role in chemistry and biology, allows control of biological interactions, affects drugs efficacy and safety, and promotes synthesis of new materials. In general, chirality manifests itself in optical activity (circular dichroism or circular birefringence). Chiral plasmonic nanoparticles have been recently developed for molecular enantiomer separation, chiral sensing and chiral photocatalysis. Here, we show that optical chirality of plasmonic nanoparticles exhibiting strong scattering can remain completely undetected using standard characterisation techniques, such as circular dichroism measurements. This phenomenon, which we term meso-chiral in analogy to meso-compounds in chemistry, is based on mutual cancellation of absorption and scattering chiral responses. As a prominent example, the meso-optical behaviour has been numerically demonstrated in multi-wound-\ch{SiO2}/Au nanoparticles over the entire visible spectral range and in other prototypical chiral nanoparticles in narrower spectral ranges. \textcolor{black}{The meso-chiral property has been experimentally verified by demonstrating chiral absorption of gold helicoid nanoparticles at the wavelength where conventional circular dichroism measurements show absence of chiral response ($g_\text{ext}$=0).} These findings demonstrate a valuable link between microscopic to macroscopic manifestations of chirality and can provide insights for interpreting a wide range of experimental results and designing chiral properties of plasmonic nanoparticles. 

\end{abstract}
 
\maketitle
\clearpage

\section{Introduction}

Chirality is an essential natural property of objects, signifying that they cannot be superimposed with their mirror image by any combination of translation and rotation operations. Its importance in chemistry and biology is elucidated by the fact that the life on the Earth itself is chirally asymmetric at a molecular level \cite{zhao2016enantioselective,zhao2017chirality}. The research in optical effects related to chirality was initially focused on sensing and spectroscopy of chiral molecules, but recently has been extended to the control of an enantiomerically pure synthesis and chiral catalysis through the utilisation of enhanced chiral responses of plasmonic nanoparticles as well as chiral properties of circularly polarised light \cite{kelly2000use,cho2023bioinspired,kim2022enantioselective,chen2024universal,ke2023vacuum,bainova2023plasmon}.  

The chiral optical response can be quantified through optical activity, represented by circular birefringence (CB) and circular dichroism (CD), related to unequal real and imaginary parts, respectively, of the permittivity of chiral media for different handedness of circularly polarised light \cite{wang2017circular, bliokh2016spin}. Due to its direct connection to absorbance/extinction spectra, circular dichroism 
$C [\text{deg}]=(S^\text{LCP}-S^\text{RCP}) \frac{\text{ln}(10)}{4}\frac{180[\text{deg}]}{\pi}$, where $S^\text{LCP}$ and $S^\text{RCP}$ are the absorbance/extinction measured for left- and right- circular polarised (LCP and RCP) light, respectively, is widely used for identification of material chirality \cite{lee2018amino}. When scattering is negligible, as often happens in the case of bulk media or molecules, absorbance and extinction \textcolor{black}{($\text{Extinction}=\text{Absorption}+\text{Scattering}$)} coincide with each other.  

Some molecules possess multiple chiral centres but do not exhibit a chiral optical signature because they have an internal plane of symmetry, as a consequence of which the optical responses of the chiral centres cancel each other. An example of a meso-compound is a tartaric acid, which has two chiral centres but is optically inactive due to its internal symmetry. Meso-compounds are important in catalysis, drug delivery and in the development of polymers \cite{hoffmann2003meso,smith2009lost,busscher20052}. 

In this letter, we show that a strongly chiral response of plasmonic nano-objects can nevertheless remain hidden in conventional optical CD measurements. \textcolor{black}{Such nanoparticles have opposite and equal chiral responses in absorption and scattering, which results in an achiral response in the measured overall extinction. In an analogy with meso-compounds in molecular chemistry exhibiting chirality cancellation between two opposite chiral centres, we term such nanoparticles meso-chiral.} The effect arises \textcolor{black}{due} to strong scattering of nanoscale objects, which is absent for molecular species. 
This property \textcolor{black}{appears} to be quite common among plasmonic nano-objects at certain wavelengths, but more surprisingly for some it can be broadband. Particularly, combining strong chirality produced by asymmetrically arranged winding \ch{SiO2} nanorods \cite{zhao2020surface} with multishell design feasible for fabrication \cite{bykov2023broadband}, a quintessential example of meso-chiral structures, multi-wound plasmonic nanoparticles (MWPNs), is demonstrated. A simulated chiral response of the MWPNs indicates the possibility of achieving a meso-chiral optical behaviour over the entire visible range. 
These findings reveal the hidden chirality which can be undetected in the standard CD spectral measurements and contribute to understanding the true chirality of nanostructures. The opposite chiral effects in absorption and scattering may have a distinct impact in a range of phenomena, including the local field enhancement effects and hot-electron generation in plasmonic nanoparticles.

\section{Results and discussion}
In contrast to molecules for which the chiral response is determined by their fixed electronic configurations, the chiral response of plasmonic nanoparticles can be engineered through their design by achieving collinear electric and magnetic dipolar resonances with the required magnitudes and phase difference at the same wavelength \cite{yuanyang-natcomm}.  
In comparison to to deeply-subwavelength-size molecules, for which extinction and absorption are practically indistinguishable as scattering is minimal, the situation maybe different for plasmonic nanoparticles with their strongly enhanced resonant scattering, even for the subwavelength sizes.  Typically, the so-called g-factor is used to represent optical chirality observed in either nanoparticle extinction, absorption, or scattering. This parameter describes the difference in nanoparticle interaction with LCP and RCP light: 
\begin{equation}
g~=~2\frac{S^\text{LCP}-S^\text{RCP}}{S^\text{LCP}+S^\text{RCP}}\label{gfactor} \, ,
\end{equation}
where $S$ is the measured signal, such as extinction, absorption or scattering intensity (according to the definition adopted in CD spectrometers, the handedness of the incident circularly polarised light is defined from the point of view of the receiver). 
The extinction g-factor is proportional to the CD value, $C$, normalised to the sum of extinction magnitudes: $ g_\text{ext}~=~\frac{4}{\text{ln}(10)} \frac{\pi}{180}\frac{2}{S^\text{LCP}_\text{ext}+S^\text{RCP}_\text{ext}}{C}$~\cite{lee2018amino}. \textcolor{black}{Using the definition of the g-factor (Eq.~\eqref{gfactor}) and the relationship between extinction, absorption and scattering signals $S_\text{ext} = S_\text{abs}+S_\text{scat}$, one can obtain that the value of $g_\text{ext}$ is always located between that of $g_\text{abs}$ and $g_\text{scat}$, as follows from the final result of the following derivation:
        \begin{equation}
        \begin{aligned}
            g_\text{ext}&=2\frac{S_\text{ext}^\text{LCP}-S_\text{ext}^\text{RCP}}{S_\text{ext}^\text{LCP}+S_\text{ext}^\text{RCP}}=2\frac{S_\text{abs}^\text{LCP}-S_\text{abs}^\text{RCP}+S_\text{scat}^\text{LCP}-S_\text{scat}^\text{RCP}}{S_\text{abs}^\text{LCP}+S_\text{abs}^\text{RCP}+S_\text{scat}^\text{LCP}+S_\text{scat}^\text{RCP}}\\&=\frac{2\sfrac{(S_\text{abs}^\text{LCP}-S_\text{abs}^\text{RCP})}{(S_\text{abs}^\text{LCP}+S_\text{abs}^\text{RCP})}\cdot\sfrac{(S_\text{abs}^\text{LCP}+S_\text{abs}^\text{RCP})}{2}+2\sfrac{(S_\text{scat}^\text{LCP}-S_\text{scat}^\text{RCP})}{(S_\text{scat}^\text{LCP}+S_\text{scat}^\text{RCP})}\cdot\sfrac{(S_\text{scat}^\text{LCP}+S_\text{scat}^\text{RCP})}{2}}{\sfrac{(S_\text{abs}^\text{LCP}+S_\text{abs}^\text{RCP})}{2}+\sfrac{(S_\text{scat}^\text{LCP}+S_\text{scat}^\text{RCP})}{2}}\\&= \frac{g_\text{abs}\overline{S_\text{abs}}+g_\text{scat}\overline{S_\text{scat}}}{\overline{S_\text{abs}}+\overline{S_\text{scat}}},
            \end{aligned}\label{gextrelationfunction}
        \end{equation}
 where $\overline{S_\text{abs}}$ ($\overline{S_\text{scat}}$) is the average of LCP and RCP absorption (scattering) intensities $\overline{S}=\sfrac{(S^\text{LCP}+S^\text{RCP})}{2}$.} This relation indicates the possibility of achieving a balance between absorption and scattering chirality values for the realisation of an achiral response in extinction.
\begin{figure}[!ht]
    \begin{center}
        \includegraphics[width=14cm]{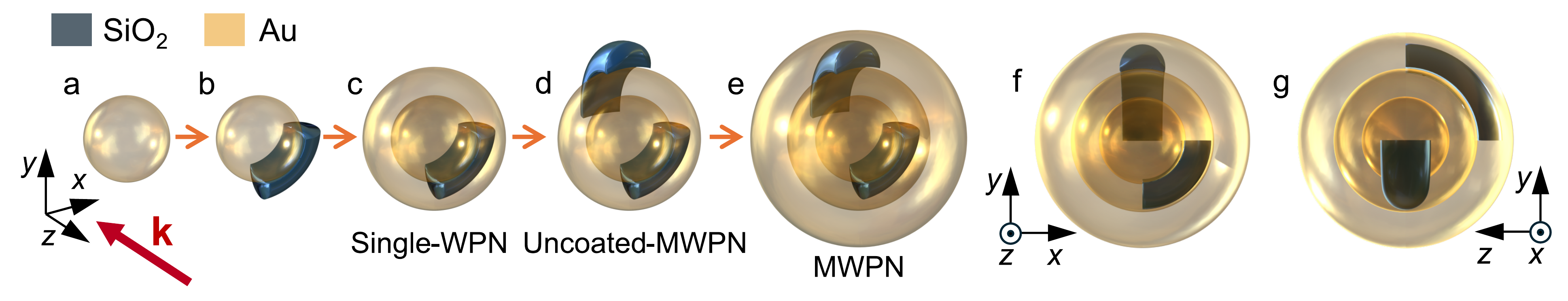}
        \caption{\textbf{MWPN fabrication process.} (a)~The MWPN is grown starting from a 60-nm gold nanosphere. (b)~A 15-nm-thick curved \ch{SiO2} nanorod of a circular shape spanning a $\pi/2$ sector with a cross-section shown in the insert, is deposited on the nanosphere. (c)~A 18-nm-thick gold shell is added. (d)~A second curved \ch{SiO2} nanorod similar to one in (b) is deposited in the plane perpendicular to the plane of the first nanorod. (e)~An outer 21-nm-thick gold shell is added. (f,g)~Cross-sections of the final nanoparticle in the (f)~$x$-$y$ and (g)~$y$-$z$ planes. It should be noted that chirality (g-factor) of the nanopaticle is only weakly dependent on the thickness of the  shells and the meso-chiral behaviour is still observed in a range of shell thicknesses.}\label{particle}
    \end{center}
\end{figure}

As an example of the meso-chiral plasmonic nanoparticles we designed multi-wound \ch{SiO2} nanorod structures on Au nanospheres, which are feasible for fabrication using layered nanorod deposition \cite{zhao2020surface,bykov2023broadband} (Fig.~\ref{particle}
). The geometrical chirality is introduced in a step by step fashion through consecutive deposition of two wound \ch{SiO2} nanorods on a plasmonic (Au) nanosphere with a encapsulation in plasmonic material after each step. The first \ch{SiO2} nanorod breaks the spherical symmetry of the structure, leaving only two symmetry planes. Then, the second \ch{SiO2} nanorod positioned in the perpendicular plane breaks the remaining symmetry, resulting in a chiral nanoparticle. 

\begin{figure}[!ht]
    \begin{center}
        \includegraphics[width=14cm]{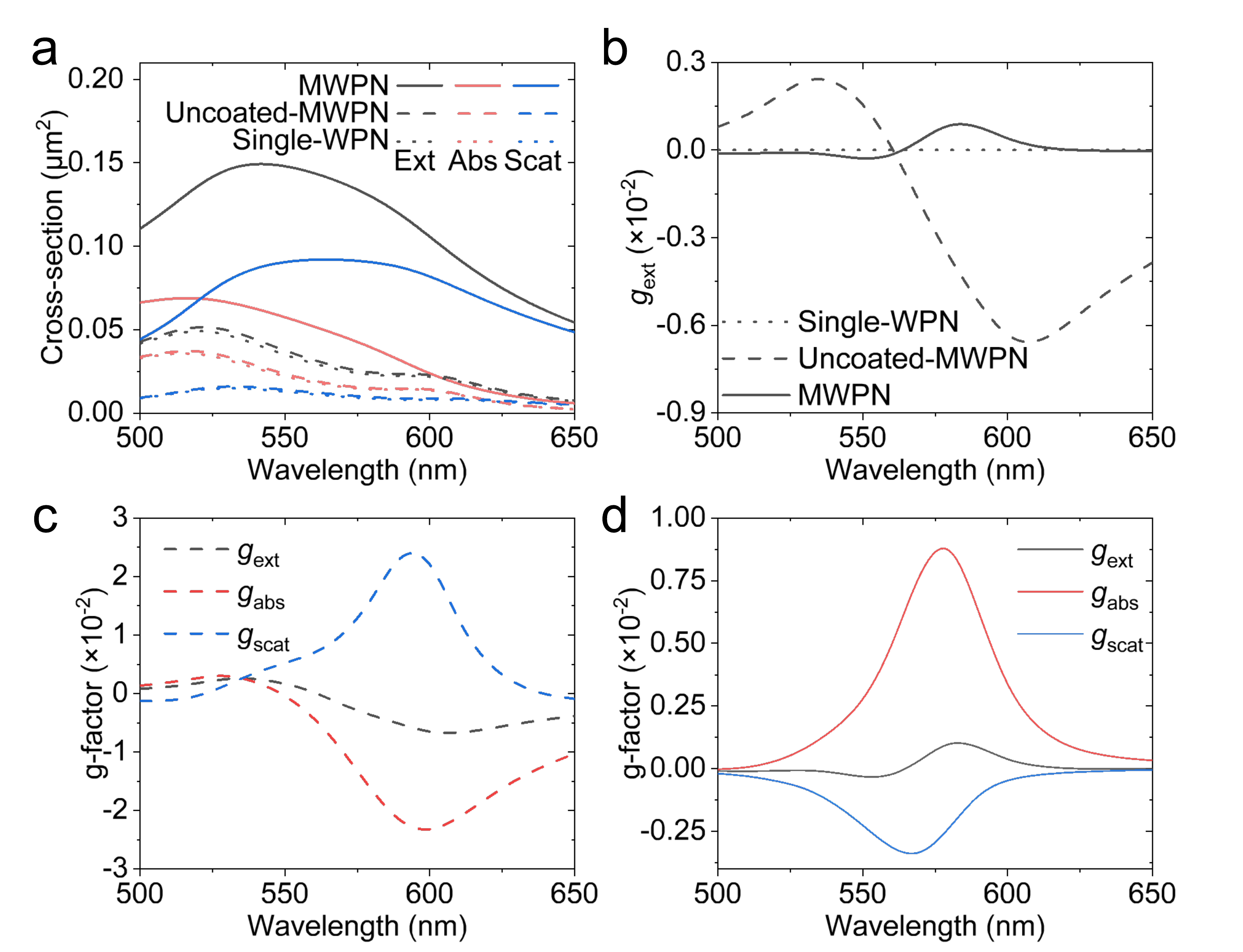}
        \caption{\textbf{Simulated optical response of MWPNs, together with corresponding spectra of g-factors.} (a)~\textcolor{black}{Simulated extinction (black), absorption (red) and scattering (blue)} cross-sections and (b) the corresponding $g_\text{ext}$ spectra of a single-WPN (dot line), an uncoated-MWPN (dash line) and a MWPN (solid line) nanostructures. (c,~d)~\textcolor{black}{Simulated $g_\text{ext}$ (black), $g_\text{abs}$ (red) and $g_\text{scat}$ (blue)} spectra of (c)~an uncoated-MWPN and (d)~a coated MWPN. The plane-wave illumination is along the $z$-direction and the parameters of the nanostructures are as in Fig.~1.}\label{MWSAspectra}
    \end{center}
\end{figure}

At all fabrication stages, the nanostructure has a pronounced resonant optical response in absorption, scattering and extinction in the visible spectral region mediated by localised surface plasmon resonances (Fig.~\ref{MWSAspectra}a). While the extinction g-factor, $g_\text{ext}$, is understandably equal to zero for a nanostructure with a single-nanorod and has a pronounced values for the uncoated MWPN, it returns to nearly-zero values in the MWPNs upon deposition of the outer Au layer (Fig.~\ref{MWSAspectra}b), even though the particle is intrinsically chiral. At the same time, uncoated MWPN and final MWPN nanoparticles possess pronounced chiral response in both scattering and absorption, which predominantly have the opposite signs (Fig.~\ref{MWSAspectra}c,d). Surprisingly, in the case of MWPNs (Fig.~\ref{MWSAspectra}d) they balance each other to produce an achiral response in extinction over the entire visible spectral range, demonstrating a broadband meso-optical behaviour. 
Thus, although these nanoparticles are fundamentally chiral, which affects not only the absorption, but the entire local field response at the nanoscale, this would remain undetected using the standard means of chiral characterisation using CD measurements.

\begin{figure}[!ht]
    \begin{center}
        \includegraphics[width=14cm]{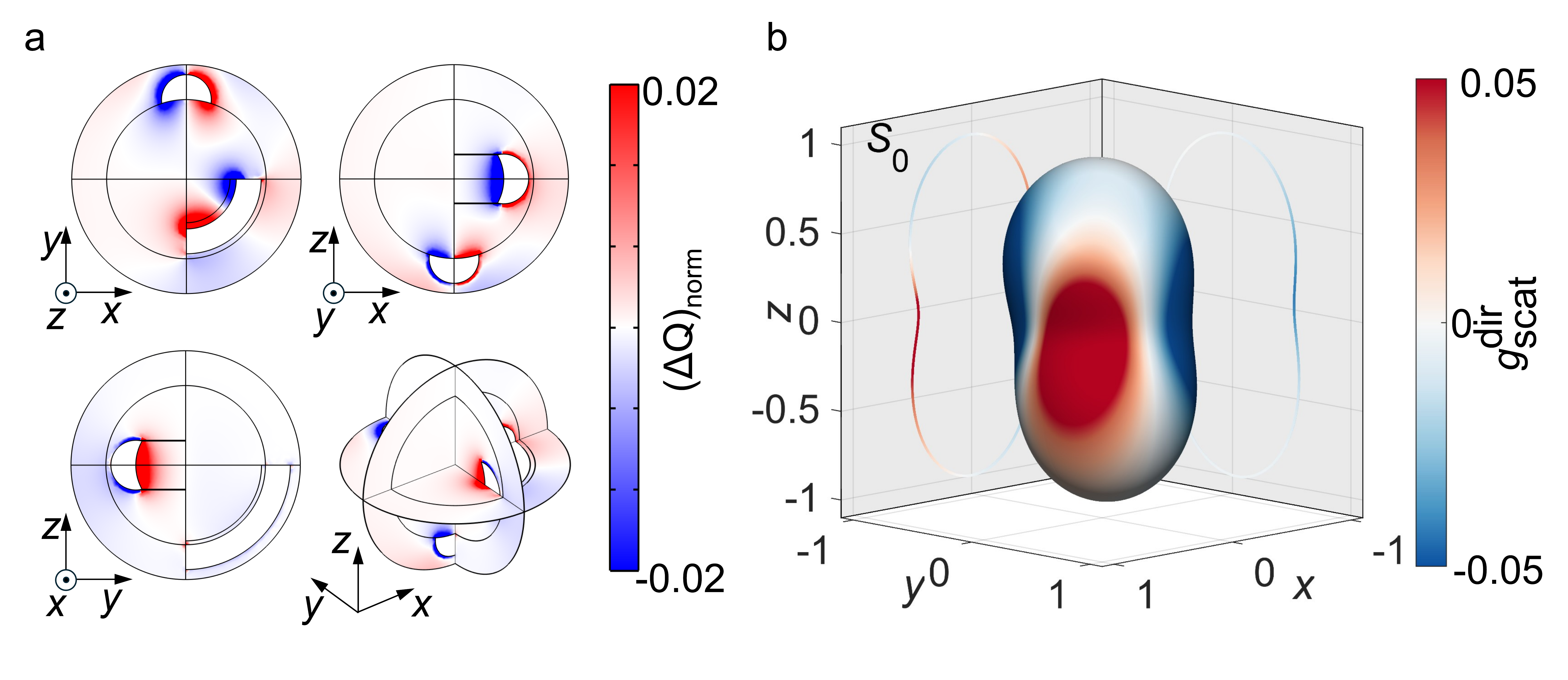}
        \caption{\textbf{Simulated spatial distributions of the difference in the optical response of MWPNs between 560-nm LCP and RCP illumination.} (a)~Cross-section maps of normalised differential absorption density $(\Delta Q)_{norm}=Q^\text{LCP}-Q^\text{RCP}$, where $\Delta Q$ is normalised by the maximum absorption value inside the nanoparticle. (b)~Normalised far-field scattering intensity (magnitude) and directional $g_\text{scat}^\text{dir}$ diagram (colour plot), where $g_\text{scat}^\text{dir}$ is the scattering g-factor measured in a given direction. The nanoparticle parameters are as in Fig.~1.}\label{field}
    \end{center}
\end{figure}
 Particularly, the second (outer) \ch{SiO2} winding rod introduces the chirality to the geometry of the nanoparticle and brings the opposite signs of $g_\text{abs}$ and $g_\text{scat}$ at the uncoated-MWPN stage (Fig.~\ref{MWSAspectra}c). In this case, $g_\text{ext}$ keeps the same sign and tendency as $g_\text{abs}$, which infers the prevailing contribution of $g_\text{abs}$ to the total optical chirality. After adding the top gold layer to produce the final MWPN structure, the signs of $g_\text{abs}$ and $g_\text{scat}$ are swapped. This addition also brings a stronger enhancement to scattering compared to that for the absorption (Fig.~\ref{MWSAspectra}a), increasing the contribution of $g_\text{scat}$ to $g_\text{ext}$ and promoting a better balance between absorption and scattering chiral responses than those in the uncoated-MWPN. Thus, the MWPN presents an intriguing example of meso-chiral properties, with a weak overall chiral response presented by $g_\text{ext}$, but prominent and opposite chirality in absorption and scattering existing in a broad wavelength range. 
 The meso-chiral property can also be inferred from the opposite signs of the difference of the magnitude of the electric field between LCP and RCP illumination, $|\mathbf{E}^\text{LCP}|-|\mathbf{E}^\text{RCP}|$, integrated inside and  outside the nanoparticles: inside the nanoparticle, the total electric field under the LCP illumination is stronger than that under the RCP illumination, while outside the situation is the opposite. 
 
 The microscopic pattern of circular dichroism is quite intricate. The enantiomeric difference in both loss and scattering varies across the nanoparticle volume and scattering directions, respectively (Fig.~\ref{field}), which can have the dependence with sign variation, upon integration returning the g-factors presented in Fig.~\ref{MWSAspectra}. In this respect, the local chirality both in absorption and scattering affects the optically-driven effects in the vicinity of the nanoparticle, while the integral values define the overall chiral response.

\begin{figure}[!b]
    \begin{center}
        \includegraphics[width=14cm]{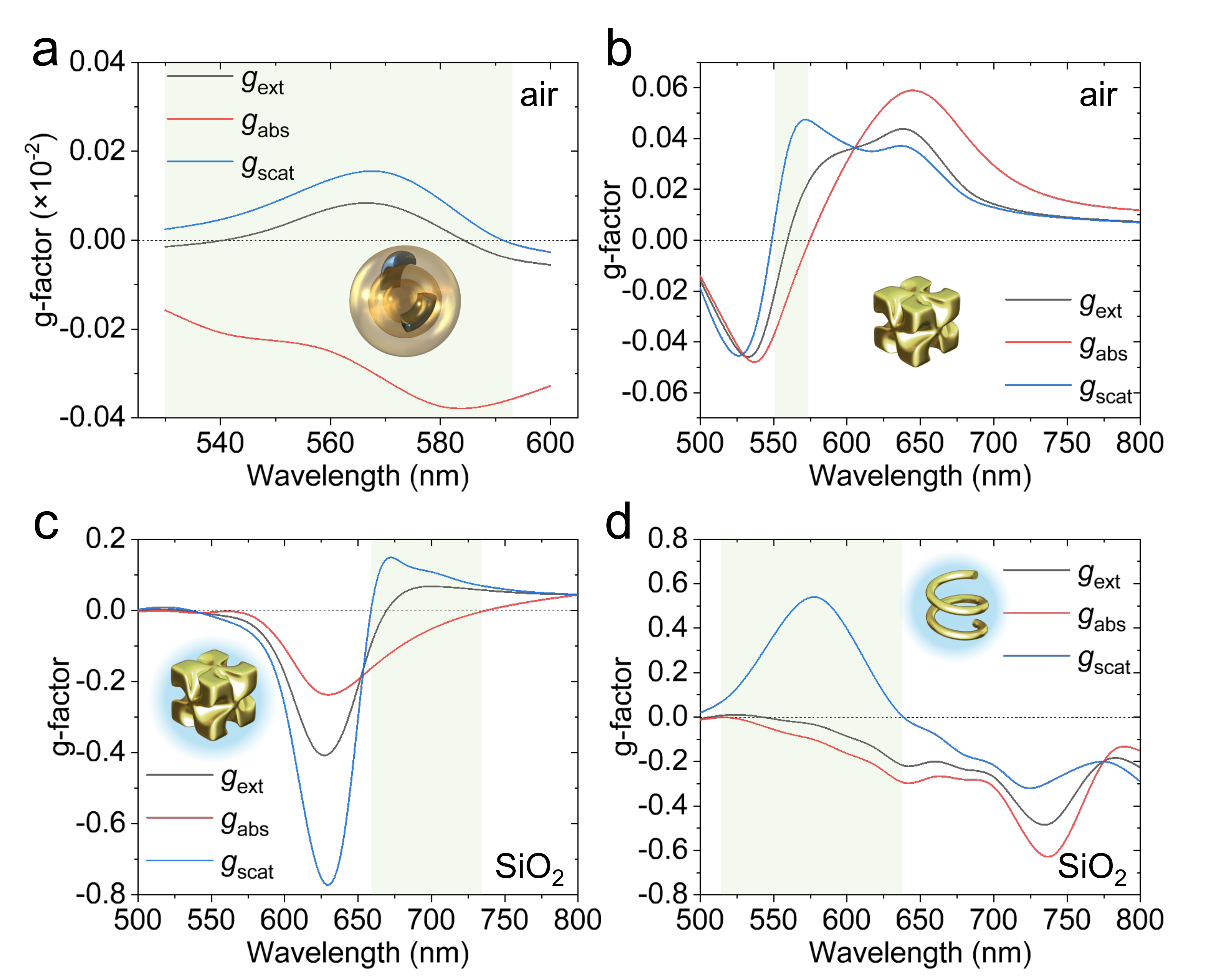}
        \caption{\textbf{Meso-chiral optical properties of common chiral plasmonic nanostructures.} Simulated g-factor spectra of (a) MWPNs averaged over 300 incident directions, (b,~c) L-helicoid with a 180-nm side length and (d) L-helix with a 180~nm major diameter, a 15~nm nanowire radius and a double twist with 105~nm period, in (a,~b) air and (c,~d) \ch{SiO2} matrix. Shading indicates the wavelength range of the meso-chiral behaviour.}\label{otherstructures}
    \end{center}
\end{figure}

\begin{figure}[!h]
    \begin{center}
        \includegraphics[width=14cm]{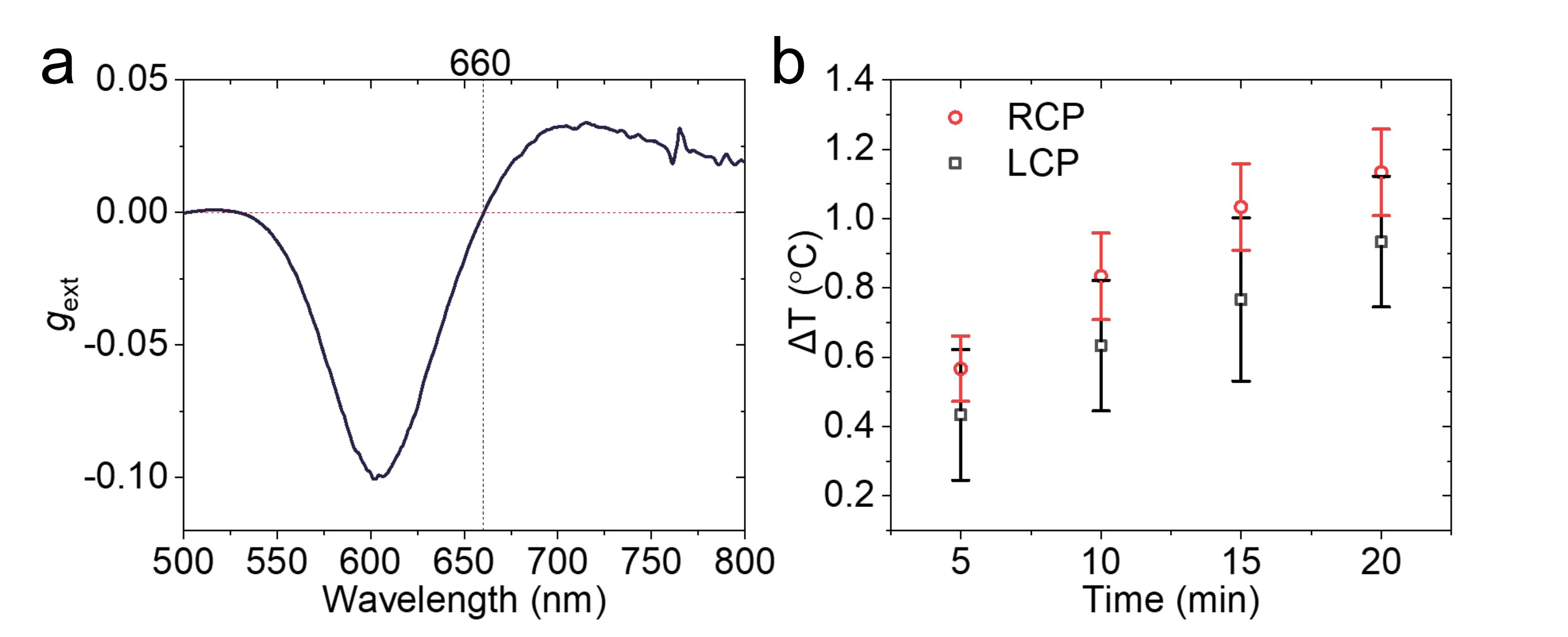}
        \caption{\textcolor{black}{\textbf{Photothermal response of helicoid nanoparticles.}}  \textcolor{black}{(a) The experimental CD spectrum for an aqueous solution of helicoid nanoparticles. A non-chiral response $g_\text{ext}$=0 is observed in extinction at the wavelength of 660~nm. (b) Optically-induced temperature changes of the helicoid solution under illumination with RCP and LCP light at a wavelength of 660 nm.}}\label{photothermal}
    \end{center}
\end{figure}
\textcolor{black}{In many chiral applications, it is important to have a reliable measure of the nanostructure chirality to correctly estimate its influence. The demonstration of a meso-chiral phenomenon calls attention to the fact that the standard way of estimation of the chiral response using the measurement of chiral extinction can be misleading: even seemingly achiral nanostructure characterised by the CD technique can affect the chiral processes through the chirality of the local fields related to its intrinsically chiral absorption and/or scattering responses. More importantly,} the introduced meso-chiral property is useful for distinguishing the contributions from the effects based on scattering and absorption in chiral nanostructures in order to choose or adjust their design for specific applications. For example, for  photocatalysis, it is the absorption cross-section is important, while scattering also plays, though, an indirect role \cite{gargiulo2019optical}. Particularly, the light absorbed by the nanoparticles generates hot carriers capable of changing the energy landscape of chemical reactions, simultaneously upon conversion to the heat increasing their speed \cite{gelle2019applications,zhang2018plasmon}. In this respect, the introduction of the absorption chirality described by $g_\text{abs}$ leads to the realisation of polarisation-sensitive photocatalysis. At the same time, the polarisation-sensitive scattering related to $g_\text{scat}$ affects the overall distribution of light over the reactor volume, also influencing photo-sensitive reactions. In this respect, meso-chiral plasmonic nanoparticles can present opposite chiral optical effects of the light scattering and the joint influence of the hot-carrier and thermal effects. Overall, observation of dissimilar photo-catalytic properties for such nanoparticles having achiral CD response would be a direct way of experimental observation of meso-chiral concept. \textcolor{black}{Also, meso-chiral particles can find their application in chiral sensing, where they, similarly to the racemic plasmonic nanoparticle arrays \cite{tan2024trace,garcia2018enantiomer}, can enhance its sensitivity, not masking a much weaker molecular signal with their own (absent) chiral response.}

While for \textcolor{black}{oriented} MWPNs (e.g., deposited on a substrate or fixed in a transparent matrix), the meso-chiral response is broadband (Fig.~2), random orientation of nanostructures as, e.g., in a colloidal solution still exhibits this property albeit in narrower spectral range (Fig.~\ref{otherstructures}a). The meso-chiral property is not so uncommon and can be observed for various geometries of plasmonic nanoparticles under different conditions. As archetypal chiral nanostructures, a gold L-helicoid \textcolor{black}{in air and a \ch{SiO2} matrix} and a gold L-helix \textcolor{black}{in air}, were numerically investigated. Notably, both of these structures can be oriented for the top illumination on a substrate or in matrix \cite{kim2022enantioselective,gansel2009gold}. The opposite signs of $g_\text{abs}$ and $g_\text{scat}$ with a weak $g_\text{ext}$ can be observed in most of the cases (apart from a helix in air), but in a narrow spectral range (Fig.~\ref{otherstructures}b-d). This indicates the possibility of hidden chirality for a wide range of plasmonic nanostructures remaining undetected in standard CD characterisation at a single wavelength and the importance of further development of techniques for optical chirality measurement.

\textcolor{black}{The meso-chiral property has been further experimentally demonstrated by studying an optical response of gold helicoid nanoparticles dispersed in water. The CD measurements reveal the meso-chiral behaviour $g_\text{ext}=0$ at a wavelength of 660 nm (Fig.~\ref{photothermal}a). Using this illumination wavelength, a photothermal response of the helicoid nanopartilces was studied, which is directly related to the absorption cross-sections. The optically-induced temperature changes were found to be significantly different for RCP and LCP light illumination, confirming the optical chirality of the nanoparticles (Fig.~\ref{photothermal}b). Thus, the nanoparticles which have a non-chiral response from circular dichroism measurements extinction, possess chiral response in absorption, which is a direct demonstration of the meso-chiral behaviour.}

\section{Conclusion}
\textcolor{black}{We have introduced a novel concept of meso-chiralilty for nanoparticles, when the substantial chirality in absorption and scattering can be hidden inside a completely achiral response in extinction due to the opposite sings of these components.} In this case, the essentially chiral local response of the nanoparticle remains undetected or significantly underestimated using \textcolor{black}{conventional} CD measurements. (It should be noted, that an optical chiral response can also be cancelled in racemic mixtures which have equal amounts of enantiomers of opposite handedness, providing the balance in the overall optical response.) At the same time, while the overall chiral response of meso-chiral nanoparticles is zero, the hidden chiral behaviour can significantly affect local enantio-sensitive light-matter interactions, e.g. polarisation-sensitive photocatalysis, holding the potential for influencing other optical processes. Notably, this property is not so uncommon in chiral plasmonic nanoparticles, which was demonstrated on the examples of multi-wound particles, helixes, and helicoids, and is experimentally verified for the latter. The hidden optical chirality and opposite chiral responses in absorption and scattering can be a powerful tool for understanding and tailoring chiral light-matter interactions.

\section*{Methods}
\subsection*{Numerical simulations}
The chiral response of multi-wound plasmonic nanoparticles was simulated using a finite element method in a scattered field formulation (COMSOL software package). The simulation domain was surrounded by a perfectly matched layer to ensure the absence of back-reflection. Apart from the nanoparticle domain in the case of the MWPN and a spherical domain closely enclosing the helicoid and helix nanoparticles, symmetrical mesh was used for all other domains to avoid potential artificial chirality created by it. The extinction cross-section $\sigma_{ext}$ was calculated as the sum of scattering ($\sigma_\text{scat}$) and absorption ($\sigma_\text{abs}$) counterparts. The latter were obtained by integrating the scattered field power flow over a surface surrounding the nanoparticles and the total field power flow entering them, respectively, followed by normalisation by the incident power flow:
\begin{equation}
        \begin{aligned}
        &\sigma_\text{abs} &&= -\frac{1}{I_0} \iint (\mathbf{n} \cdot \mathbf{P}_\text{tot}) dS, & \quad \text{(a)}\\
        &\sigma_\text{scat} &&= \frac{1}{I_0} \iint (\mathbf{n} \cdot \mathbf{P}_\text{scat}) dS, & \quad \text{(b)}
        \end{aligned}\label{extabsscateq}
        \end{equation}
where $I_0$ is the intensity of the incident light, $\mathbf{n}$ is the vector normal to the nanoparticle surface, $\mathbf{P}_\text{tot}$ and $\mathbf{P}_\text{scat}$ are the Poynting vectors of the total fields and scattering fields, respectively. The g-factors were calculated from the relative difference of the cross-sections under the LCP and RCP illumination (Eq.~\eqref{gfactor}). For nanoparticles preferentially oriented on a substrate or in matrix, plane-wave illumination under normal incidence (along the $z$-axis) was considered.

\textcolor{black}{For randomly oriented nanoparticles, averaged results from 300 nanoparticle orientations were used to simulate the case of colloidal systems. Particularly, the orientation of the nanoparticle was fixed, while illumination directions were set using Fibonacci approach ~\cite{Fibonaccimethod}: 
\begin{equation}
        \left. \begin{aligned}
        &\theta_\text{j} &&= \arccos (1-2j/N)\\
        &\phi_\text{j} &&= 2j\pi \Phi^{-1}
        \end{aligned} \right\}(0\leq j < N)  \, , \label{extabsscateq}
        \end{equation}
where $\theta_j$ and $\phi_j$ are the polar and azimuthal illumination angles respectively, $N$ is the number of sample directions and $\Phi$ is the Fibonacci golden ratio $(1+\sqrt{5})/2$.
The g-factors were evaluated as
\begin{equation}
        \begin{aligned}
        g_{i}=~2\frac{\sum S_{i}^\text{LCP}-\sum S_{i}^\text{RCP}}{\sum S_{i}^\text{LCP}+\sum S_{i}^\text{RCP}} \, , \label{avgangle}
        \end{aligned}
        \end{equation}
which represents the definition of the g-factor, where $\sum S_i$ expresses the summation of either extinction, scattering or absorption signals $S$ over the sampled illumination directions.}

\subsection*{\textcolor{black}{Experimental characterisation}}
\textcolor{black}{The helicoid particles were fabricated by the method described in Ref.~\cite{yuanyang-natcomm}. Briefly, the helicoids were grown for two hours at $30~^\circ$C from 320~$\mu$L octahedral seeds dispersed in a solution containing 29.04~mL DI water, 6.4~mL 0.1~M Hexadecyltrimethylammonium bromide (CTAB) and 640~$\upmu$L 10~mM \ch{HAuCl4}, 3.6~mL 0.1~M ascorbic acid and 16~$\upmu$L 5~$\upmu$M L-glutathione. Then, the particles ware rinsed by centrifuging (2000$\times g$, 3~min) 3~times and kept in a 1~mM CTAB solution for further use.}

\textcolor{black}{CD spectra of the nanoparticles in water were measured using commercial CD spectrometer (Applied Photophysics Chirascan Plus). The photothermal measurements were performed with collimated white light from a supercontinuum laser (NKT Photonics SuperK-EVO-HP), filtered at the wavelength of 660~nm$\pm$10~nm, circularly polasrised using a linear polariser and a quarter waveplate. The temperature of a 1~mL helicoid solution with 200~rpm stirring in a 1-cm-thick glass cuvette was measured by a digital thermometer (RS PRO Handheld Digital Thermometer 206-3738). To check the consistency of the results, the experimental measurements were repeated 3 times. The stability of the nanoparticles under laser excitation was checked by performing CD measurements before and after the illumination. }

\section{References}
\bibliographystyle{IEEEtran}
\bibliography{bib}
\section*{Acknowledgement}
The authors are grateful to Dr.Tam Bui and Optical Spectroscopy Facility for CD spectra measurements.

\section*{Research funding}

This work was supported by the UK EPSRC project (EP/W017075/1). Y.X. acknowledges support from China Scholarship Council. 

\section*{Author contributions}
Y.X., and A.V.Z. developed the concept, Y.X. and A.V.K. performed the numerical simulations, Y.X performed the measurements, all authors contributed to the writing of the manuscript.

\end{document}